# Possible Giant Orbital Paramagnetism in Nanometer Scale 2DEG Strips

Michael J. Harrison

*Department of Physics and Astronomy, Michigan State University, East Lansing, MI 48824-2320, USA*

**Abstract.** An elementary calculation shows that Landau diamagnetism becomes significantly altered and very large paramagnetic effects emerge at low temperature in nanoscale 2DEG strips penetrated by a perpendicular applied magnetic field and bounded by a parabolic potential, such as may arise from negative voltage applied to a split gate. These novel results are described by an expression which manifests the total system magnetization as a difference between evolved orbital paramagnetism and altered diamagnetism. These predicted effects correspond to drift motion of electrons parallel to the strip length arising from Landau eigenstates that are non-degenerate in the combined presence of a perpendicular applied magnetic field and electric fields associated with a confining parabolic potential. A new heterostructured magnetic material based on orbital electronic motion in 2DEG strips is proposed.

## INTRODUCTION

The Landau diamagnetism of a confined system of free electrons [1] has been the starting point for investigations by a number of workers [2-13] with the view of obtaining corrections that arise from the presence of a bounding surface. The correction terms may be paramagnetic or diamagnetic, and depend sensitively on the increase in potential energy at the surface relative to the Fermi energy [2]. The role of surface boundary conditions may be expected to increase as the size of the system decreases and the surface to volume, or perimeter to area, ratio increases. Measurements on small metal particles have revealed size-dependent paramagnetic enhancement at low temperatures [14]. Experiments have also shown that the orbital magnetism of small isolated Ga-As heterostructures in the shape of squares several times $10^3$ nm on a side exhibit paramagnetic orbital contributions 100 times larger than the magnitude of the Landau diamagnetic susceptibility [15]. Subsequently a number of theoretical papers have appeared [16-20] which have carefully explored the question of orbital magnetization in both two-dimensional systems and crystalline solids.

## DIMINISHED ROLES OF PARTICLE INTERACTION AND DISORDER SCATTERING

There have been numerous experimental and theoretical studies of the roles of disorder and interparticle interactions in affecting charge motion in 2D electron systems [21, 22, 23, 24, 25, 26]. Ballistic transport in heterostructures has been achieved in 2DEG constrictions with widths up to 250 nm for samples that have quasiparticle mean free paths as long as 30 μm [27, 28]. We surmise that the underlying ideas of Fermi-liquid theory [29] subtend our understanding of ballistic quasielectron motion and long mean free paths observed in certain experimental studies of reduced dimensionality electron systems, typified by those cited above. These experiments indicate that heterostructures can now be fabricated in which independent quasiparticles move freely over paths long compared to tens of microns [28, 30]. We shall therefore stipulate that the nanometer scale 2DEG strips studied in this paper contain independent particles that move with little interaction or scattering in the presence of magnetic fields over times much longer than quasiparticle cyclotron periods and over mean free paths greater than typical sample dimensions.

We calculate in this paper the orbital magnetism of narrow 2DEG strips bounded laterally by a parabolic potential energy of the sort that may be approximately produced between the arms of a split-gate electrode [30, 31] maintained at a negative potential V. The lateral electron confinement which defines the strip width in our calculation then arises from a potential energy that is a maximum on the split-gate electrode and vanishes on the center line bisecting the 2DEG strips along their lengths. The electronic systems we consider range in width upwards from 125 nm, and extend along their length indefinitely. We further assume temperatures sufficiently low to adopt a zero temperature approximation, and areal electron densities $n_0$ commonly achieved in two-dimensional heterostructures, ie. $n_0 \approx$ several times $10^{11}$ or $10^{12}$ cm$^{-2}$. The new expression obtained for the magnetisation per particle in units of the effective Bohr magneton for a 2DEG strip of width $2L_y$ and areal electron density $n_0$ is

$$M(B,K) = [(2/3)(\pi L_y^2 n_0)(Kb^{-3})(1+K/b^2)^{-2} - (1+K/b^2)^{-1/2}] , \qquad (1)$$

where $b \equiv 2B/(n_0\varphi_0)$ is a dimensionless magnetic field associated with a magnetic field B in gauss, $\varphi_0 \equiv hc/|e|$ is the magnetic flux quantum, and K is a parameter defined below related to the strength of the parabolic potential which confines the 2DEG strip.

The importance of the new result embodied in Eq.(1) is that it manifestly separates the paramagnetic consequences of the parabolic potential in which the 2DEG moves from the resulting diamagnetism which is seen to weaken relative to the Landau result of (-1) per electron as the dimensionless magnetic field b diminishes because either B becomes small or the areal density $n_0$ becomes large.

## MODEL 2DEG STRIP IN PERPENDICULAR MAGNETIC FIELD

Consider a 2DEG strip in the x-y plane running parallel to the x-axis, which bisects the strip. A magnetic field **B** points in the positive z direction. A quadratic confining potential energy U(y) varies in the y direction within the region $-L_y \leq y \leq L_y$, where $2L_y$ represents the width of the 2DEG strip. With the Landau gauge choice of **A** = ( -By, 0, 0 ) for the vector potential and U(y) = $\gamma^2 m y^2/2$ as the confining parabolic potential we have the one-electron Hamiltonian

$$H = [(p_x - By|e|/c)^2/2m] + [p_y^2/2m] + [\gamma^2 m y^2/2] , \qquad (2)$$

where e = -|e| is the electronic charge, m is the quasielectron effective mass, and $\gamma^2$ measures the strength of the confining transverse parabolic potential energy from the negative voltage on the confining split-gate electrode.

For periodic boundary conditions over a length $L_x$ in the x direction the eigenfunctions of this Hamilitonian are known to be [32]

$$\Psi_{n,k}(x,y) = ( e^{ikx}/\sqrt{L_x} ) \, u_n(y-Y) , \qquad (3)$$

where $u_n(y)$ is a normalized harmonic oscillator wave function for level number n, $Y = Y(k) = \hbar k \Omega/\omega^2 m$, $\Omega = |e|B/mc$ is the cyclotron frequency, and $\omega^2 = \Omega^2 + \gamma^2$. Because $-L_y \leq Y \leq L_y$ within the strip, the quantum number k is restricted to the range $-k_{max} \leq k \leq k_{max} \equiv \omega^2 m L_y/\hbar\Omega$.

The energy eigenvalues to which the eigenfunctions given by Eq.(3) correspond are [32]

$$\varepsilon_{nk} = \hbar\omega (n + \tfrac{1}{2}) + (\gamma^2/\omega^2)(\hbar^2 k^2/2m) . \qquad (4)$$

The capacity to fabricate high electron mobility heterojunctions and nanoscale lithography have made it possible to create an electrostatically squeezed 2DEG [30] whose quasielectrons have mean free paths exceeding 10 μm at low temperatures. Electron-impurity mean free paths as long as 30 μm have been realized [28]. We shall therefore assume that for the magnetic fields and nanoscale strip widths under consideration, the electrons with energy levels given by Eq.(4) move without collisions at the low temperatures we stipulate. And we take $2N_0$ such electrons to occupy the area $2L_x L_y$.

The emergence of paramagnetism with the presence of an attractive transverse parabolic potential may be understood by considering the way energy levels change with magnetic field. When $\gamma^2 = 0$ and B>0 the usual Landau levels are obtained, and for the simplest case with n=0 one has for these lowest levels $\varepsilon_{0k} = \hbar\Omega/2$, so that $\partial\varepsilon_{0k}/\partial B > 0$ results in diamagnetism since the energy levels rise above their zero field values, so that the magnetization *M* is negative: $M \sim -\partial\varepsilon_{0k}/\partial B < 0$. But for $\gamma^2 > 0$ and B >0 one finds from Eq.(4) that whenever k is bounded according to $(m/2\hbar\gamma^2)\omega^3 < k^2 < k_{max}^2$ we have $\partial\varepsilon_{0k}/\partial B < 0$, so that $M \sim -\partial\varepsilon_{0k}/\partial B > 0$, and contributions towards paramagnetism occur. By summing up the contributions of all occupied states by integration the net magnetism of the system is then obtained.

The allowed separation of adjacent k values is $\delta k = 2\pi/L_x$ so that the allowed number of k values for electrons within the area $2L_x L_y$ is $2k_{max}/\delta k$, and the total number of states including spin, D, in any Landau band n and associated with the area $2L_x L_y$ is $D = 4k_{max}/\delta k = (2\omega^2 m L_x L_y)/(\pi\hbar\Omega)$. It is instructive to write D as

$$D = 4 (BL_x L_y/\varphi_0) [1 + (\hbar\gamma/(2\mu B))^2 ] , \qquad (5)$$

where $\mu \equiv |e|\hbar/(2mc)$ is the effective Bohr magneton corresponding to effective mass m. It is to be noted that D is greater than the total number of flux lines threading the area $2L_x L_y$ by a factor which depends on the strength of the confining parabolic potential energy compared to the electron spin Zeeman energy in the magnetic field.

The spread of energy eigenvalues, Q, constituting the Landau band width for any given quantum number n is given by

$$(\varepsilon_{n,k_{max}} - \varepsilon_{n,0}) = m\omega^2\gamma^2 L_y^2/(2\Omega^2) \equiv Q \ . \tag{6}$$

Since the energy level dependence on k is through $k^2$, and the states with negative k, $-k_{max} \leq k < 0$, have the same distribution of energies as the positive k states, $0 \leq k \leq k_{max}$, the total density of states within Landau band n for electrons within the area $2L_xL_y$ is twice that which arises from states with $0 \leq k \leq k_{max}$ and $0 \leq Y \leq L_y$. Therefor, including spin, the total density of states for Landau band n is then given by $\rho_n(\varepsilon) = [2(2/\delta k)]/|\partial\varepsilon_{nk}/\partial k|$, which is then calculated for n=0 to be

$$\rho_0(\varepsilon) = [D/(2Q^{1/2})] \, [\varepsilon - \hbar\omega/2]^{-1/2} \tag{7}$$

for $(\hbar\omega/2) \leq \varepsilon \leq (\hbar\omega/2) + Q$. It is convenient to define a dimensionless magnetic field b given by $b \equiv 2B/(\varphi_0 n_0) = 2|e|B/(hcn_0)$. Then with the Fermi energy of a non-interacting 2DEG of areal density $n_0$ and effective mass m given by $E_F = \pi\hbar^2 n_0/m$ we define a parameter K that expresses the strength of the confining parabolic potential energy by

$$K \equiv (\hbar\gamma/E_F)^2 \ . \tag{8}$$

The number of states D in a Landau band may now be written as

$$D = 2N_0 b \, [1 + K/b^2] \ , \tag{9}$$

which illustrates directly how low values of dimensionless field b may nevertheless result in large number of states in a Landau band when a confining parabolic potential energy is present.

The use of a parabolic potential to represent the electronic confining potential in the 2DEG region between the arms of a split-gate at negative voltage only approximates the quantum well created by the split-gate [33, 34]. It neglects both screening and possible effects of charged defects that may have entered the system. But we anticipate that for the narrow channels under consideration the parabolic potential adequately represents the qualitative features of quasielectron confinement over most of the area between the arms of the split-gate, and we expect that it describes the essential physics of the system.

We can readily establish an approximate relationship that connects the dimensionless parameter K with the negative statvoltage $V_{ST} < 0$ applied to the arms of the split-gate electrode that confines the quasielectrons in the 2DEG strip. If we regard the two arms of the split-gate as sufficiently long, then their potentials separately satisfy LaPlace's equation in cylindrical coordinates in the region $|y| < L_y$ between them, if we neglect screening. Accordingly, by superposition the net potential from the two arms for points $|y| < L_y$ between them may be written in the form

$$\xi(y) = A \, [\log(L_y - |y|) + \log(L_y + |y|)] \quad , \tag{10}$$

where A is a constant to be determined. To order $y^2$ we may write $\xi(y) \approx A\log L_y^2 - Ay^2/L_y^2$, and dropping the constant term while taking $A = -V_{ST}$ we have $\xi(y) \approx V_{ST}(y^2/L_y^2)$ for $|y| \leq L_y$. Comparison of this approximate potential $\xi(y)$ with the parabolic potential energy in Eq. (2) shows that $\gamma^2 \approx -2|e| V/(300mL_y^2)$, where $V < 0$ is the negative mks voltage on the split-gate electrode. From the expression for $E_F$ and that for K in Eq.(8) we also obtain $\gamma^2 = \pi^2\hbar^2 n_0 K/m$. Setting these two expressions for $\gamma^2$ equal to each other we obtain the following approximate relation between V and K:

$$-V = (300\, \pi^2\hbar^2 n_0^2 L_y^2)\, K/(2|e|m) \text{ volts} . \tag{11}$$

The Fermi level of the electronic system at $T = 0^0$ Kelvin, $\eta(0)$, when all 2DEG quasielectrons are in the lowest Landau band may be calculated from

$$\int_{\hbar\omega/2}^{\eta(0)} \rho_0(\varepsilon)\, d\varepsilon = 2 N_0 , \tag{12}$$

where $\rho_0(\varepsilon)$ is given by Eq.(7) incorporating the definitions of D and Q.

The integration in Eq.(12) is readily performed and the result is solved for the Fermi level in the upper limit:

$$\eta(0) = (\hbar\omega/2) + Q(2N_0/D)^2 . \tag{13}$$

At $T = 0^0$ Kelvin the Helmholtz free energy coincides with the energy E(0) which is given by

$$E(0) = \int_{\hbar\omega/2}^{\eta(0)} \varepsilon\, \rho_0(\varepsilon)\, d\varepsilon . \tag{14}$$

Upon integration and substitution of Eq.(13) for the Fermi level the result obtained is

$$E(0) = 2 N_0 [ (Q/3)(2N_0/D)^2 + (\hbar\omega/2) ] . \tag{15}$$

The magnetization $\mathcal{M}$ of the system is given by the derivative of E(0) with respect to magnetic field B, times (-1):

$$-\mathcal{M} = \partial E(0)/\partial B . \tag{16}$$

By carrying out the differentiation in Eq.(16), and noting the dependence on B of all quantities entering E(0) in Eq.(15) we find that subsequent algebraic manipulation and collection of terms gives the result:

$$\mathcal{M}(B) = \mu(2N_0) \, M(B,K) \, , \qquad (17)$$

where M(B,K) is given by Eq.(1) and we recall that the dimensionless field b which enters M(B,K) is given above for B in gauss. For B measured in Tesla the expression for b in Eq.(1) becomes multiplied by a factor of $10^4$.

The function M(B,K) depends sensitively on $L_y$, $n_0$, and particularly on K, which is related to the split-gate voltage through Eq.(11). In Figures.(1-6) we have plotted graphs of M(B,K) versus magnetic field in Tesla for two values of $L_y$, 125 nm and 215 nm, and two specified values of areal density $n_0$, $10^{11}$ cm$^{-2}$ and $5 \times 10^{11}$ cm$^{-2}$, for various values of confining split-gate voltage. In all six examples the values of the average magnetization per particle exceeds the magnitude of the Landau diamagnetic moment for an unbounded 2DEG by an order of magnitude or more. And the magnetic fields required are well within achievable laboratory values.

In order to have the 2DEG quasielectrons in a magnetic field contained within the strip width $2L_y$, $|Y| < L_y$, the magnetic fields must also be strong enough for the wave functions $u_n(y-Y)$ to decrease exponentially over distances, $\Delta y_0$, significantly shorter than $L_y$. An upper bound to $\Delta y_0 < L_y$ is given by the exponential decay length of the harmonic oscillator wave functions [35] for an electron in a magnetic field, $\Delta y_0 = \sqrt{[2\hbar c/(|e|B)]}$. For B=$10^3$ gauss, $\Delta y_0$=115 nm, and we have accordingly restricted the magnetic fields in Figures 1-6 to be greater than $10^3$ gauss for $L_y \geq 125$ nm.

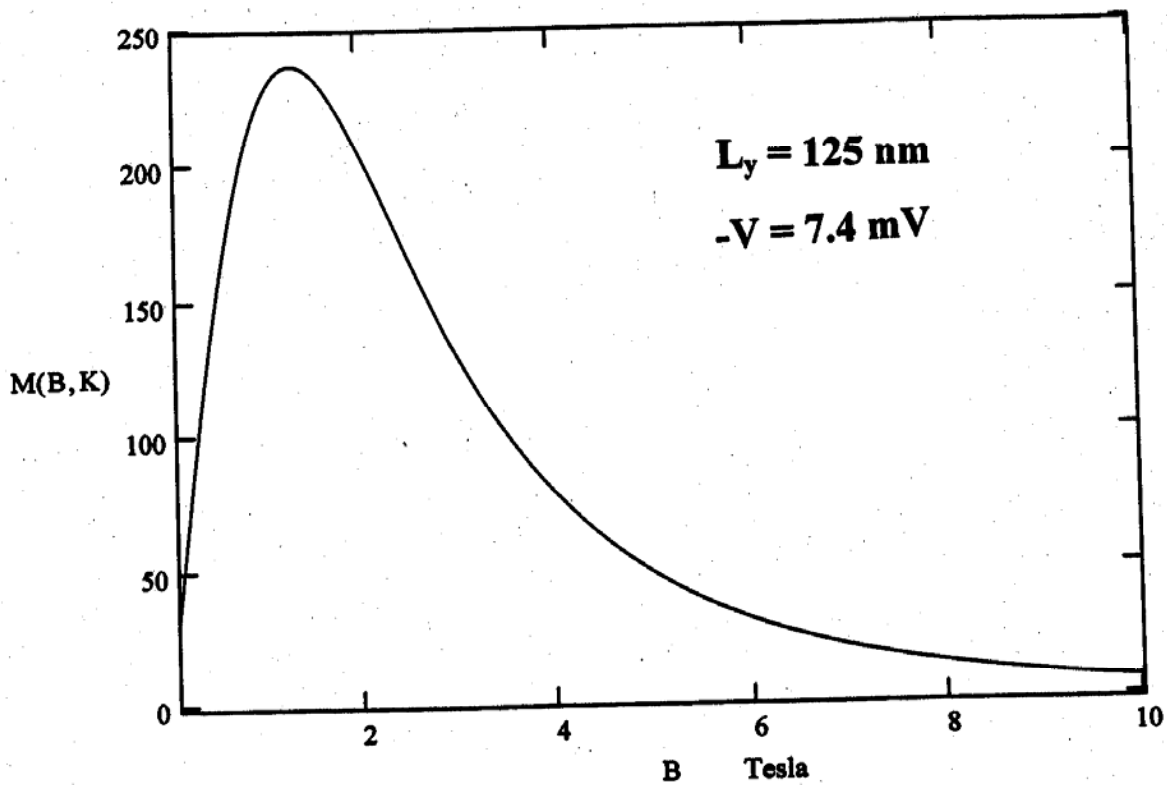

**FIGURE 1.** Magnetisation per particle in units of effective Bohr magnetons as a function of field in Tesla, for a strip of width 250 nm, areal density 5x $10^{11}$ $cm^{-2}$, and K=0.05, corresponding to $-V$=7.4 mV.

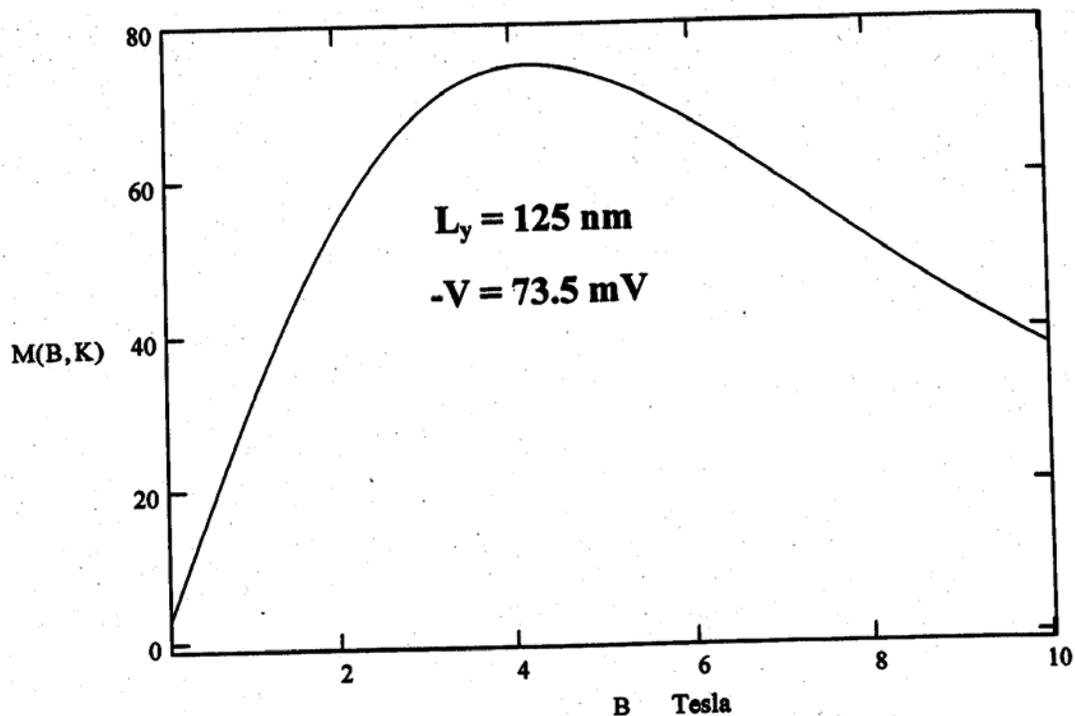
**FIGURE 2.** Magnetisation per particle in units of effective Bohr magnetons as a function of field in Tesla, for a strip of width 250 nm, areal density 5x $10^{11}$ cm$^{-2}$, and K=0.5, corresponding to $-V$=73.5 mV.

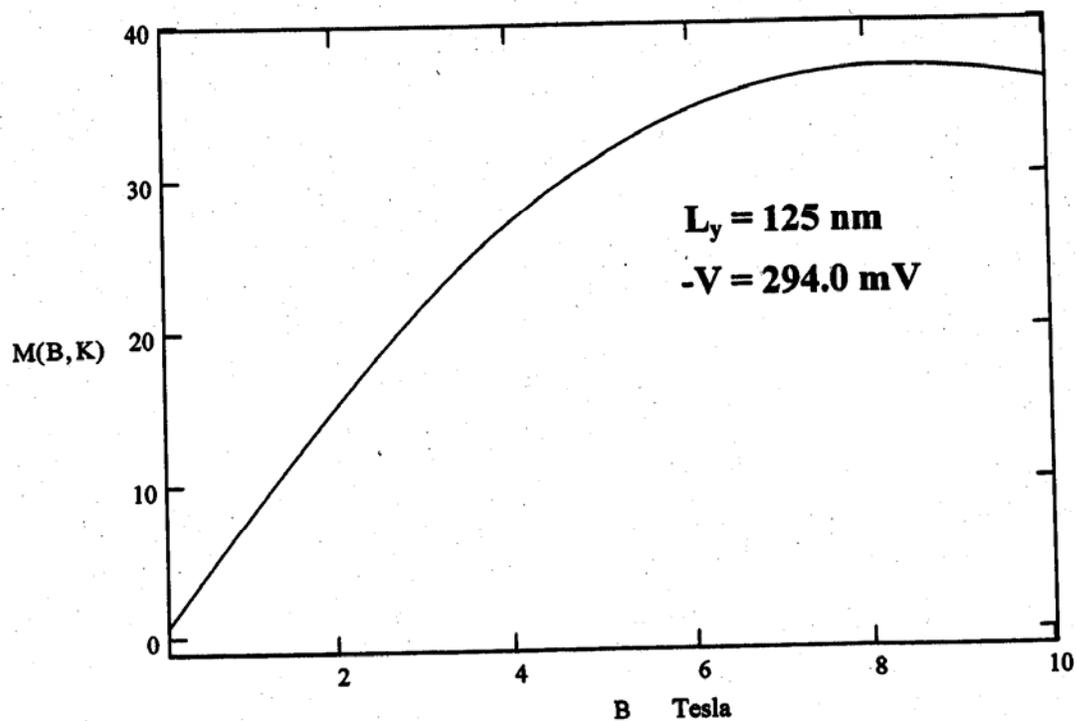
**FIGURE 3.** Magnetisation per particle in units of effective Bohr magnetons as a function of field in Tesla, for a strip of width 250 nm, areal density 5x $10^{11}$ cm$^{-2}$, and K=2.0, corresponding to $-V$=294.0 mV.

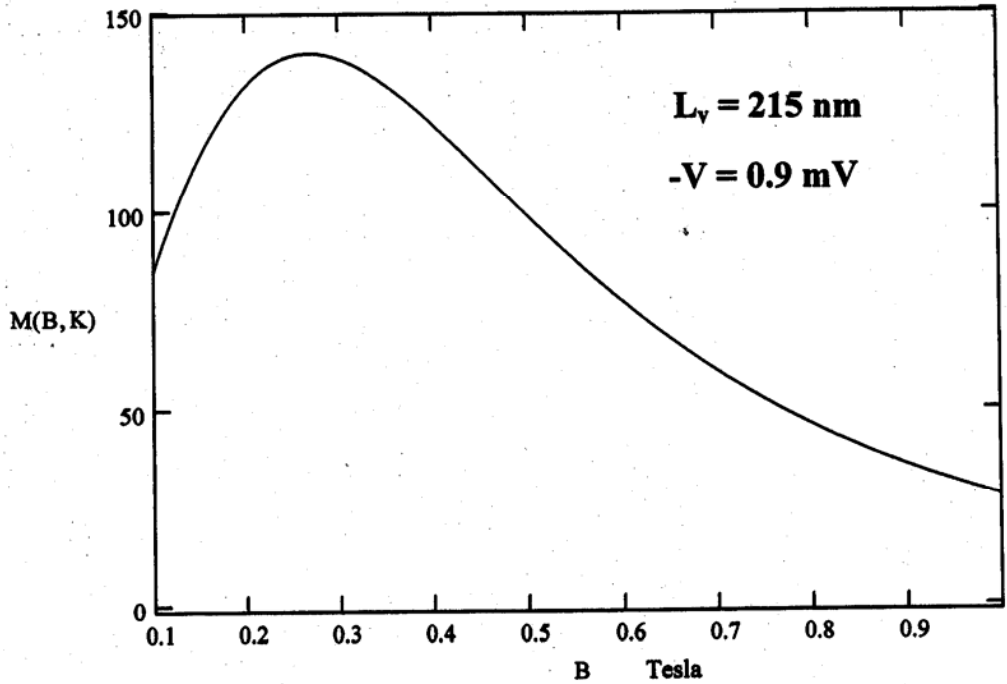

**FIGURE 4.** Magnetisation per particle in units of effective Bohr magnetons, as a function of field in Tesla for a strip of width 430 nm, areal density $1 \times 10^{11}$ cm$^{-2}$, and K=0.05, corresponding to $-V=0.9$ mV.

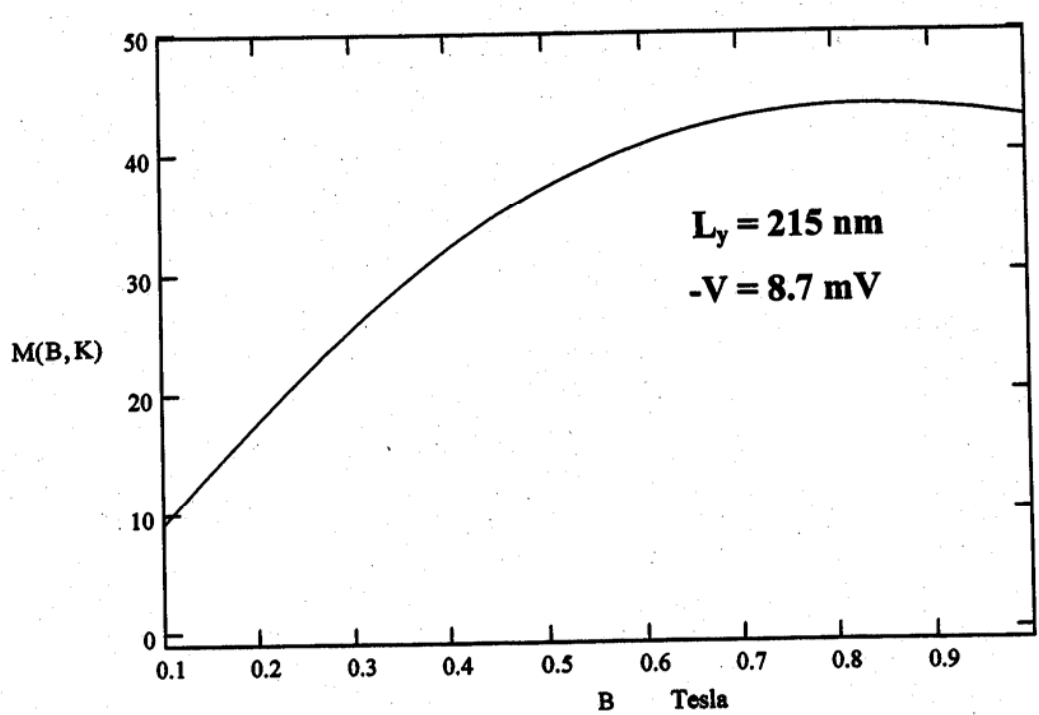

**FIGURE 5.** Magnetisation per particle in units of effective Bohr magnetons, as a function of field in Tesla for a strip of width 430 nm, areal density 1x $10^{11}$ cm$^{-2}$, and K=0.5, corresponding to −V=8.7 mV.

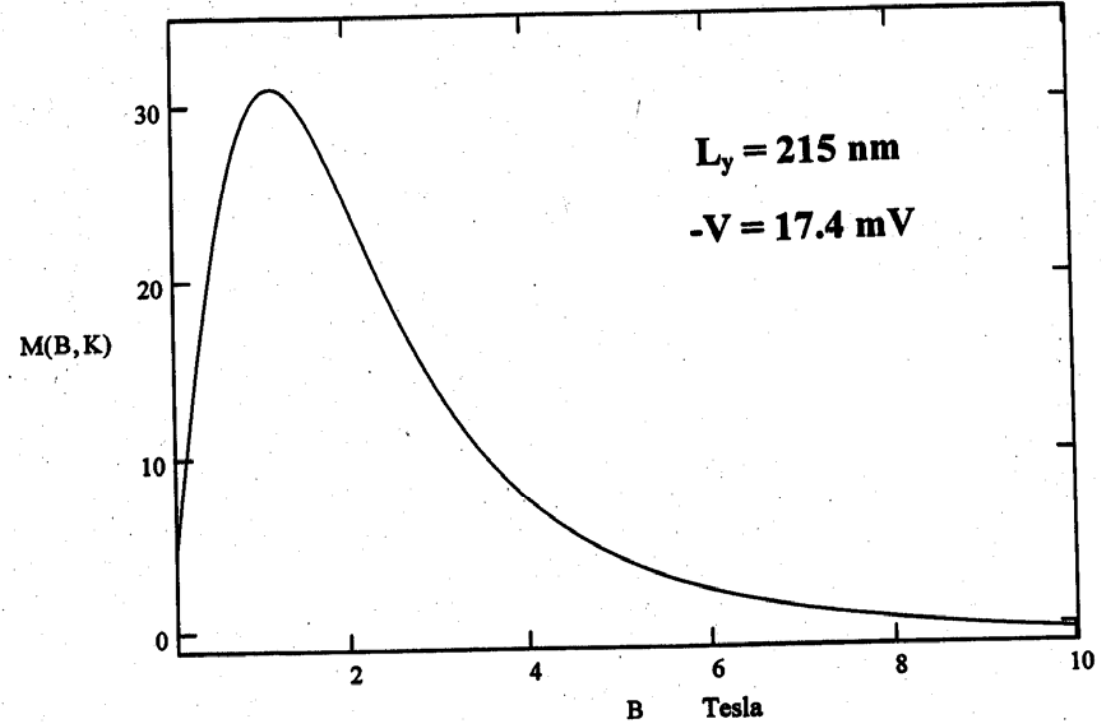

**FIGURE 6.** Magnetisation per particle in units of effective Bohr magnetons, as a function of field in Tesla for a strip of width 430 nm, areal density 1x $10^{11}$ cm$^{-2}$, and K=1.0, corresponding to −V=17.4 mV.

The large paramagnetic response of the 2DEG strips in a magnetic field may be understood semiclassically by considering the drift motion of the electrons in the quantum states they occupy. For an electron whose eigenfunction is centered at $Y = \hbar k\Omega/(m\omega^2)$, one has k>0 for Y>0 and k<0 for Y<0. But in an eigenstate whose energy is given by Eq.(4) the electron of charge e = - |e| drifts along the x direction with a velocity

$$v = \hbar^{-1} d\varepsilon_{nk}/dk = (\hbar\gamma^2/m\omega^2) k . \qquad (18)$$

From Eq.(18) one deduces that the drift is towards +x for Y>0 and towards −x for Y<0, with $|v| = (\gamma^2/\Omega)|Y|$ attaining its maximum value near the strip edges where $|Y| \approx L_y$. Since electrons have negative charge their electric currents are oppositely directed to their drift velocity, and the greatest currents come from near the edges of the strip where |Y| is greatest. The right-hand rule relating current direction and the direction of magnetic fields produced shows that <u>within</u> the 2DEG the magnetic fields produced by the drifting electrons are in the <u>same</u> direction as the applied magnetic field B.

## POSSIBLE NEW MATERIAL

It may be possible to construct an heterostructure which is strongly paramagnetic based on the orbital paramagnetism of confined 2DEG strips discusses above. Consider a set of confined 2DEG strips arranged in a plane parallel to each other and uniformly separated by a distance α . Construct a stack together of similar such planes containing confined parallel 2DEG strips, parallel to each other and located periodically with interplane perpendicular separation l.. Then the volume magnetization density of such a system in the presence of a magnetic field perpendicular to the stack is given by $|\mathbf{m}| = (f/l)n_0\mu M(B,K)$, where $f = (2L_y/\alpha)$ is the fraction of a plane covered by 2DEG strips.

## ACKNOWLEDGMENTS
I would like to thank Professor Walter Kohn for calling my attention to the 1953 paper of Frank S. Ham, Professor Morrel H. Cohen and Professor S. D. Mahanti for their invaluable criticism and helpful comments.

## REFERENCES

1. L. Landau, Z. Physik **64**, 629 (1930)
2. F.S. Ham, Phys. Rev **92**(5) , 1113 (1953)
3. R.V. Denton, Z. Physik **265**, 119 (1973)
4. F. Meier and P. Wyder, Phys. Rev. Lett. **30**(5) , 181(1973)
5. M. Robnik, J. Phys. A: Math Gen**19**, 3619 (1986)
6. J.M. van Ruitenbeck, Z. Phys. D **19**, 247 (1991)
7. B.L. Altshuler, Y. Gefen, and Y. Imry, Phys.Rev Lett **66**, 88 (1991)
8. S. Oh, A.Yu. Zyuzin, and R.A. Serota, Phys.Rev.B **44**(16), 8858 (1991)
9. B.L. Altshuler, Y. Gefen, Y. Imry, and G. Montambaux, Phys.Rev.B **47**(16), 10335 (1993)
10. R.A. Serota and A.Yu. Zyuzin, Phys.Rev.B **47**(11), 6399 (1993)
11. F. von Oppen, Phys.Rev.B **50**(23) , 17151 (1994)
12. D. Ullmo, K. Richter, and R.A. Jalabert, Phys.Rev.Lett.**74**(3), 383 (1995)
13. J-X Zhu and Z.D. Wang, Phys. Lett. A **203**, 144 (1995)
14. K. Kimura and S. Bandow, Phys.Rev.Lett.**58**(13), 1359 (1987)
15. L.P. Levy, D.H. Reich, and L. Pfeiffer & K. West, Physica B **189**, 204 (1993)
16. G. Vignale, Phys.Rev.Lett. **67**(3), 358 (1991)
17. Di Xiao, et al,Phys.Rev.Lett **95**(13), 137204-1 (2005)
18. T. Thonhauser,et al, Phys.Rev.Lett. **95**(13), 137205-1 (2005)
19. Davide Ceresoli, et al, Phys.Rev.B **74**(2), 024408-1 (2006)
20. Junren Shi, et al, Phys.Rev.Lett. **99**(19), 197202-1 (2007)
21. B. J. van Wees, et al, Phys.Rev.Lett **60**(9), 848 (1988)
22. B. Brill and M. Heiblum, Phys.Rev.B **54**(24), R17 280 (1996)
23. S. Q. Murphy, et al, Phys.Rev.B **52**(20), 14 825 (1995-II)
24. O. E. Dial, et al, Nature **448,** 176 (2007)
25. W. R. Clarke, et al, Nature Phys.**4**, 55 (2008)
26. O. E. Dial, et al, Nature **448,** 176 (2007)
27. C. Beenakker and H. van Houten; *Solid State Physics,* **44**, 110 (1991)
28. R. N. Gurzhi, A. N. Kalinenko, and A. I. Kopeliovich, Phys.Rev.B **52**(7), 4744 (1995-I)
29. L. D. Landau, Sov.Phys. JETP; **3**, 920 (1957); **5**, 101 (1957), **8**, 70 (1959)
30. N. K. Patel, et al, Phys.: Condens Matter **2**, 7247 (1990)
31. H. Buhmann, et al, Physica E **6**(1-4), 310 (2000)

32. M.J. Harrison, Phys.Rev.B **45**(7), 3815 (1992).
33. K.-F. Berggren, et al, Phys.Rev.Lett.**57**(14), 1769 (1986)
34. D.A. Wharam, , et al, Phys.Rev.B, **39**(9), 6283 (1989-II)
35. L.I. Schiff, *Quantum Mechanics – Second Edition,* McGraw-Hill Book Co., 61 (1955)